\documentclass[12pt]{iopart}

\usepackage{graphicx}
\usepackage{xcolor}
\usepackage{pdfpages}

\begin{document}

\title{Sub 20 meV Schottky barriers in metal/MoTe$_2$ junctions}

\author{Nicola J Townsend$^{1,+}$, Iddo Amit$^{1,+}$, Monica F Craciun$^2$ and Saverio Russo$^1$}

\address{$^1$ University of Exeter, School of Physics, College of Engineering, Mathematics and Physical Sciences, Exeter, EX4 4QL, United Kingdom}
\address{$^2$ University of Exeter, School of Engineering, College of Engineering, Mathematics and Physical Sciences, Exeter, EX4 4QF, United Kingdom}
\address{$^+$ these authors contributed equally to this work}

\ead{\mailto{s.russo@exeter.ac.uk}}

\begin{abstract}
The newly emerging class of atomically-thin materials has shown a high potential for the realisation of novel electronic and optoelectronic components. Amongst this family, semiconducting transition metal dichalcogenides (TMDCs) are of particular interest. While their band gaps are compatible with those of conventional solid state devices, they present a wide range of exciting new properties that is bound to become a crucial ingredient in the future of electronics. To utilise these properties for the prospect of electronics in general, and long-wavelength-based photodetectors in particular, the Schottky barriers formed upon contact with a metal and the contact resistance that arises at these interfaces have to be measured and controlled.  We present experimental evidence for the formation of Schottky barriers as low as 10 meV between MoTe$_2$ and metal electrodes. By varying the electrode work functions, we demonstrate that Fermi level pinning due to metal induced gap states at the interfaces occurs at 0.14 eV above the valence band maximum. In this configuration, thermionic emission is observed for the first time at temperatures between 40 K and 75 K. Finally, we discuss the ability to tune the barrier height using a gate electrode.
\end{abstract}

\submitto{\TDM}

\maketitle

\section{Introduction}
The emerging class of atomically thin materials has captured the interest of the research community due to the versatile physical phenomena that they exhibit\cite{wang2012electronics,jariwala2014emerging}. To name a few, the indirect to direct band gap transition as a function of layer number in atomically thin semiconductors\cite{Splendiani2010}, gate-controlled modulation of the band structure in few layer graphene\cite{craciun2009trilayer} and phase structure transitions\cite{Cho2015} hold the key to future innovations in electronic devices. Within this broad spectrum of materials, semiconducting transition metal dichalcogenides (TMDCs) are a major focal point for the wide scientific community working on fundamental and applied aspects of device physics due to their energy gaps of 1-2 eV.\cite{kang2013band,gong2013band} These gap values are ideally suited for electronics and optoelectronics, making TMDCs the prime candidates to replace bulk semiconductors in applications where added functionality, such as mechanical flexibility, is required. Atomically thin semiconductors also have the potential to readily form previously hard to access energy barriers, mostly in the few meV regime.\cite{vabbina2015highly} This feature stems from the wide variation in properties (\textit{e.g.} work function and band gap values\cite{kang2013band,gong2013band}) found within the layered materials family. As such, by tailoring the energy barrier formed at the contacts to specific applications, TMDCs can be extremely valuable for light detection in the far infrared regime as internal photoemission diodes,\cite{Lao2014} replacing other structures that are costly, or difficult to fabricate by any other means. 

One approach to realise low energy barriers is by forming a Schottky junction between a TMDC and a metal. Several recent studies on few-layer TMDCs, such as MoS$_2$, WSe$_2$ and MoTe$_2$ have identified Schottky barriers as the primary contributors to the observed contact resistance.\cite{Guo2015b,Cho2015,Allain2015,Kwon2017,Lin2014a,Pradhan2014,Fathipour2014} Indeed, Guo \textit{et al}. showed that metal induced gap states (MIGS) are the primary cause for barrier formation and that the Schottky barrier height (SBH) cannot be significantly altered by changing the work function of the contact metal.\cite{Guo2015b} In that work the formation of the Schottky barrier has been attributed mainly to the lack of dangling bonds normal to the crystallographic plane of TMDCs. The self terminating plane means that strong bonds between metal electrodes and the semiconductor cannot be easily formed, a fact that contributes to the increased contact resistance.\cite{Allain2015} Experimentally, SBH values of 80 meV were reported with the thermionic emission process being dominant down to 100 K for MoTe$_2$\cite{Lin2014a,Pradhan2014} and 60 K for MoS$_2$.\cite{Kwon2017} However, lower SBH values have, so far, not been achieved at cryogenic temperatures.  At temperatures between 200 and 400 $^{\circ}C$, a barrier height of 23 meV has been reported where the MoTe$_2$ flake is transferred onto pre-defined gold contacts, which may be due to a strain induced structural phase change from semiconducting 2H phase to a metallic 1T' phase at the contacts.\cite{Qi2017} 

Here we report the formation of ultra-low effective SBHs down to 10 meV on MoTe$_2$/metal structures, that is aided by Fermi level pinning at the interfaces. MoTe$_2$ is a TMDC with three different structural phases, the semiconducting 2H, and the semimetalic T$_{\scriptsize \textrm{d}}$ and 1T' phases. The 2H phase consists of three hexagonal planes in a ``sandwich'' formation, with Te at the outer planes and Mo at the centre. These planes are covalently bonded in a trigonal prismatic configuration that constitutes a single layer, and the layers are held together by van der Waals forces.\cite{Dawson1987a,Ruppert2014} The observed ultra-low effective SBHs in MoTe$_2$ play a minor role in the conduction at room temperature, but become significant at cryogenic temperatures and manifest in a strongly rectifying behaviour. By analysing the current-voltage characteristics of the conducting channel at different temperatures and various gate biases, we are able to extract its SBH and conductivity values. Within these systems, we find that thermionic emission persists as the dominant mechanism of transport at temperatures that range between 40 K and 75 K, whereas at higher temperatures (T $\geq$ 80 K) other transport mechanisms become more prominent. By varying the electrode material, we show that MIGS pin the Fermi level at the interface at a level of 0.14 eV over the valence band maximum. Finally, we briefly discuss the modulation of the effective SBH with applied gate bias. Our results pave the way for the realisation of far infrared devices in TMDCs and provide insight on the mechanisms of transport in MoTe$_2$ at cryogenic temperatures.

\section{Methods}
Two terminal field effect devices, schematically shown in figure \ref{fig1}a were fabricated by mechanically exfoliating MoTe$_2$ flakes from a synthetic crystal (\textit{HQ Graphene}) and transferring them onto highly doped silicon substrate with a high quality thermally grown oxide layer. The Si/SiO$_2$ serves as a global gate electrode and gate dielectric, respectively. Metal contacts were patterned using a regular electron beam lithography procedure immediately prior to metallisation. The devices were then thermally annealed for 2 hours in a H$_2$/Ar environment at 200 $^\circ$C.

MoTe$_2$ flakes were characterised using Raman spectroscopy on a Renishaw Raman microspectrometer using a 532 nm laser. Atomic force microscopy (AFM) micrographs were obtained on an Innova AFM system (\textit{Bruker Inc.}) working in the ``tapping mode'', using a Nanosensors high-reflectivity probes operating at a resonance frequency of $\sim 320$ kHz, with a nominal radius of curvature of 10 nm or smaller. Once the quality of the flakes were established, the sample was loaded into a helium-3 cryostat, where the source-drain current \textit{vs} source-drain voltage (I$_{{\scriptsize \textrm{ds}}}$-V$_{{\scriptsize \textrm{ds}}}$) response curves and the source-drain current \textit{vs} gate-source voltage (I$_{{\scriptsize \textrm{ds}}}$-V$_{{\scriptsize \textrm{gs}}}$) were recorded at decreasing temperatures between 80 K and 40 K using a home-built electrical characterisation setup.

\section{Results and Discussion}

\begin{figure}[htb!]
\includegraphics[width=\linewidth]{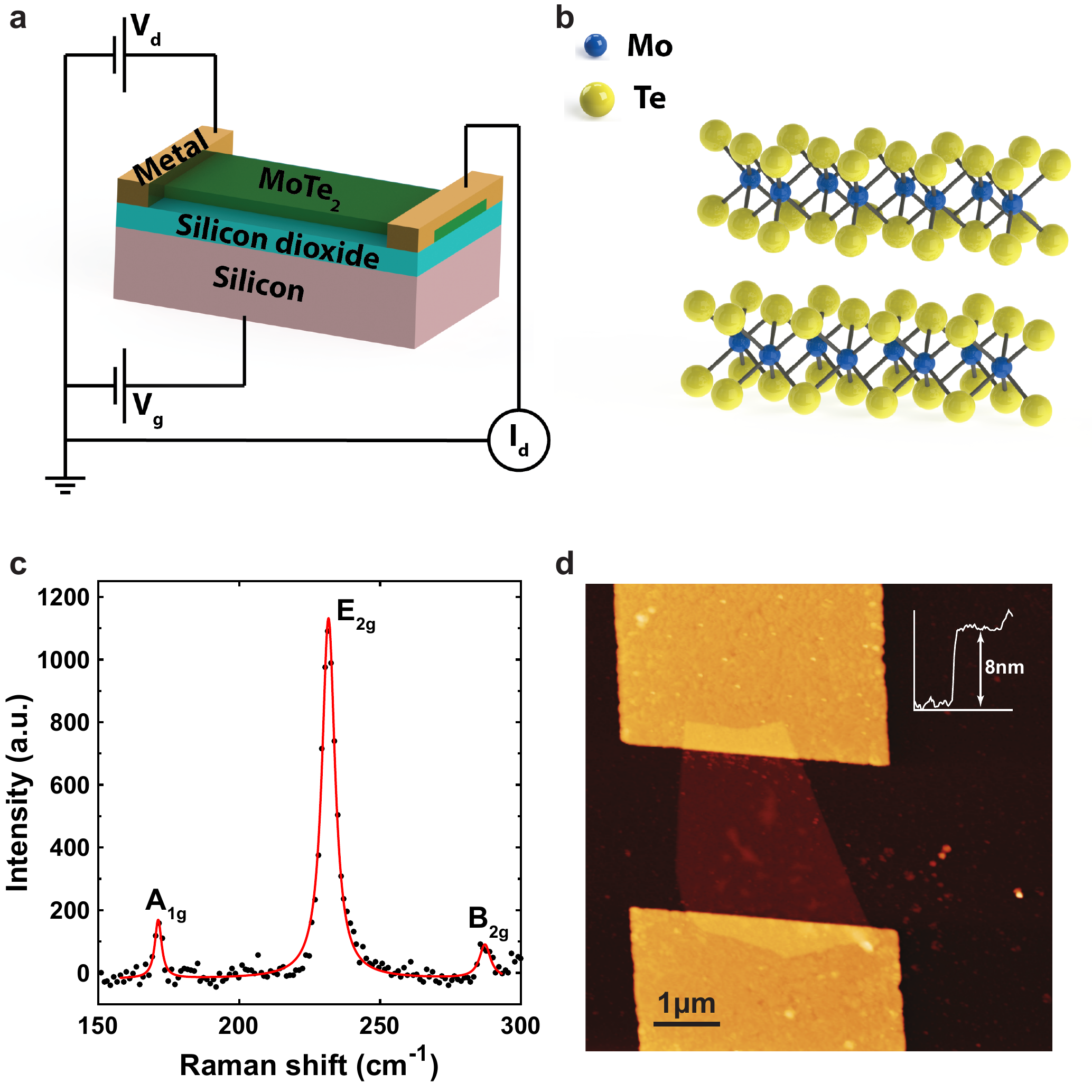}
\caption{\textbf{Material characterisation} \\ \textbf{a} Schematic of an FET device using an MoTe$_2$ flake as the channel. The drain is the biased electrode and the current is measured at the source contact.  The silicon layer acts as a global gate electrode with the silicon dioxide acting as the gate dielectric. \textbf{b} Crystal structure of 2H-MoTe$_2$ showing tellerium atoms (in yellow) covalently bonded to the molybdenum atoms (blue) in a ``sandwich'' formation. The covalently bound layers are held together by van der Waals forces. \textbf{c} Raman spectrum of typical flake showing $A_{\scriptsize \textrm{1g}}$, $E_{\scriptsize \textrm{2g}}$ and $B_{\scriptsize \textrm{2g}}$ peaks associated with few layer, 2H-MoTe$_2$. \textbf{d} AFM micrograph of a typical device with the height profile in the inset showing the flake has a thickness of 8 nm, equivalent to around 10 layers.}
\label{fig1}
\end{figure}

The devices, shown schematically in figure \ref{fig1}a, were characterised using Raman spectroscopy in ambient conditions to ensure that the flakes crystalline structure is the 2H phase (see figure \ref{fig1}b). The three peaks in the Raman spectrum (figure \ref{fig1}c), at $\sim170$ cm$^{-1}$, $\sim230$ cm$^{-1}$ and $\sim290$ cm$^{-1}$ are the finger print modes of the 2H phase, namely the $A_{{\scriptsize \textrm{1g}}}$, $E_{{\scriptsize \textrm{2g}}}$ and $B_{{\scriptsize \textrm{2g}}}$ modes respectively.\cite{Chang,Kan2015,Yamamoto2014} The $B_{\scriptsize{\textrm{g}}}$ ($\sim163$ cm$^{-1}$) and $A_{\scriptsize\textrm{g}}$ ($\sim260$ cm$^{-1}$) peaks, which are associated with the 1T' phase\cite{Kan2015} are absent indicating that the crystalline phase is indeed the semiconducting trigonal structure.  As the T$_{\scriptsize \textrm{d}}$ phase does not exist in room temperature, its Raman peaks are not considered here.\cite{Thomsen2016a} 

The number of layers was determined from the thickness of the flake as measured using atomic force microscopy (AFM). The device topography and structure are shown in the AFM micrograph (figure \ref{fig1}d). The cross section profile in the inset of the figure shows that the thickness of the flake is 8 nm, corresponding to 10 layers.\cite{Yamamoto2014}  All the flakes used in this work were in the range of 5-10 layers.

\begin{figure}[htb!]
\includegraphics[width=\linewidth]{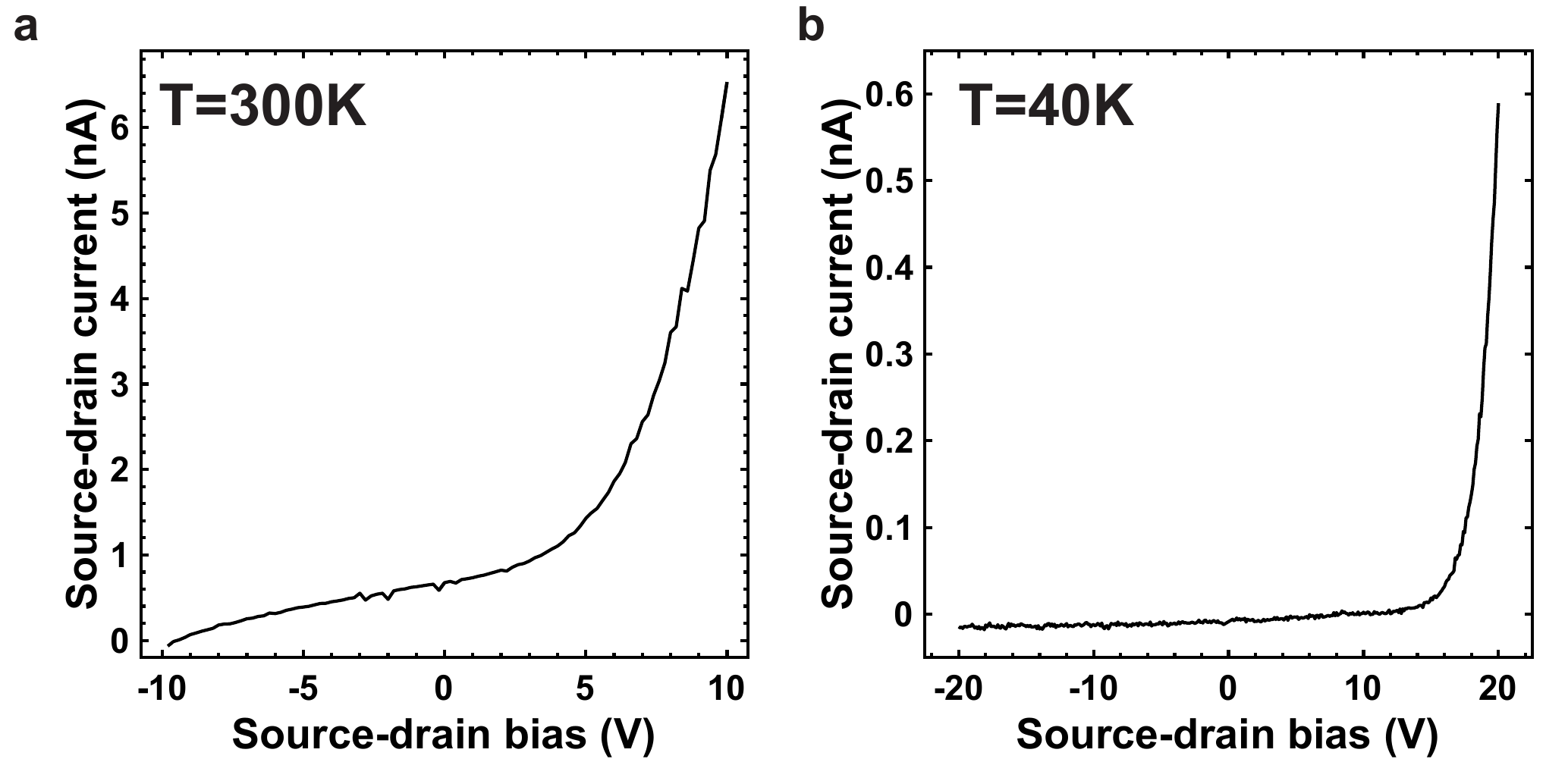}
\caption{\textbf{Response curves at different temperatures}\\
\textbf{a} The response (I$_{{\scriptsize \textrm{ds}}}$-V$_{{\scriptsize \textrm{ds}}}$) curve for a device with Au contacts taken at room temperature while the device is in the ``open'' state at V$_{{\scriptsize \textrm{gs}}} = -40$ V. \textbf{b} The response curve for the same device taken at 40 K while the rest of the parameters are constant.}
\label{fig2a}
\end{figure}

The few-layer MoTe$_2$ FET bears two identical metal contacts and should present current response characteristics that are symmetric about the zero source-drain bias (V$_{\scriptsize \textrm{ds}} = 0$ V). However, the response curve (I$_{\scriptsize \textrm{ds}}$-V$_{\scriptsize \textrm{ds}}$) shown in figure \ref{fig2a}a is sub-linear and slightly rectifying, a trend that becomes more pronounced at lower temperatures (figure \ref{fig2a}b). The increasingly rectifying response to V$_{\scriptsize \textrm{ds}}$ with decreasing temperature is indicative of the formation of asymmetric Schottky barriers at the interfaces between the metal contacts and the semiconducting channel (see supplementary information section II online).  While the semiconductor interface potential at the grounded (source) electrode is pinned by the bottom gate, and is effectively a fixed energy barrier, the electrostatic potential of the biased (drain) interface decreases the barrier height with forward bias.\cite{Tian2014}

\begin{figure}[htb!]
\includegraphics[width=\linewidth]{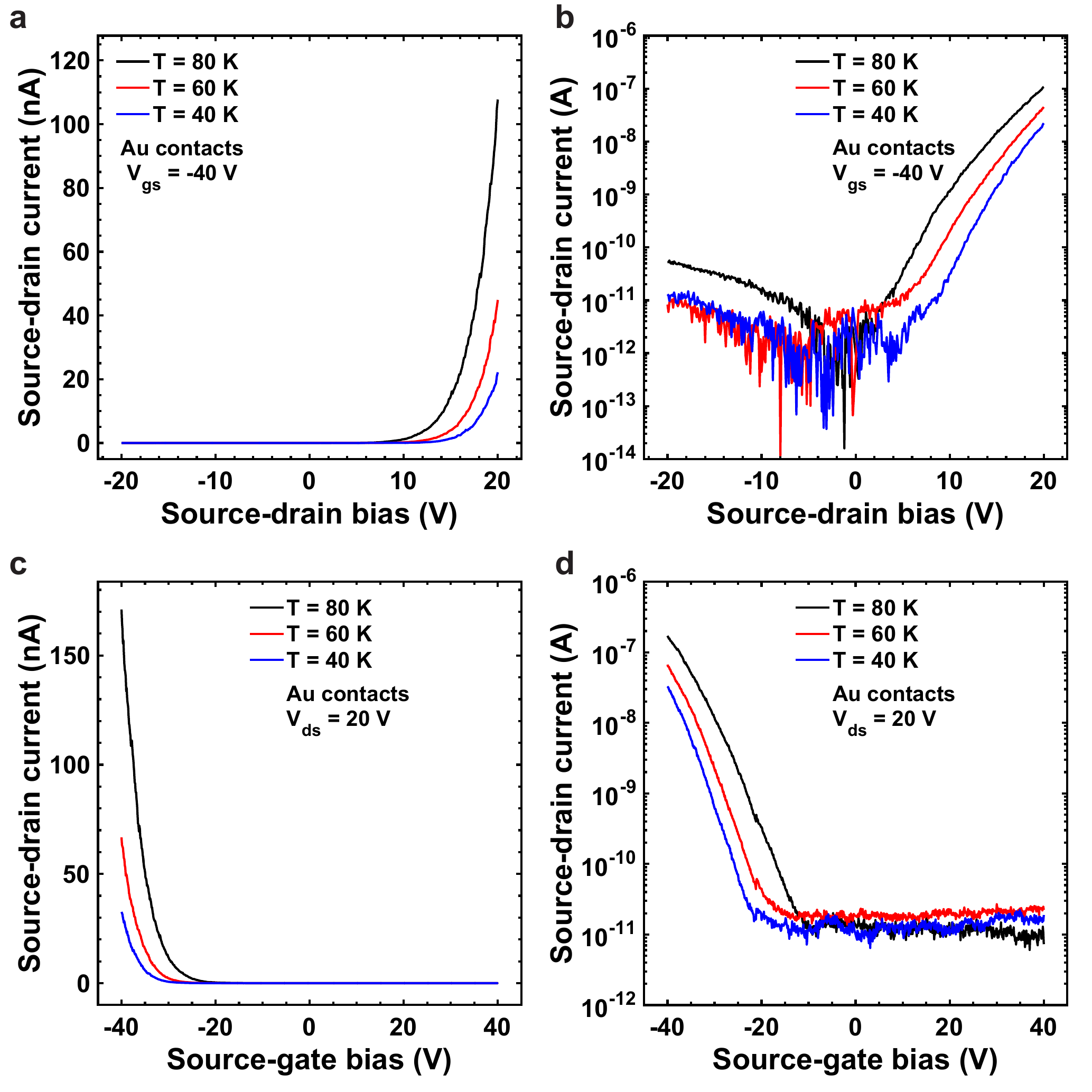}
\caption{\textbf{Electrical characterisation}\\
\textbf{a} Response curves for a device with Au contacts at temperatures between 80 K and 40 K, measured with the devices in the  ``open'' state at V$_{\scriptsize \textrm{gs}}=-40$ V. The curves exhibit a diode-like rectifying characteristics, as is further emphasised in the semi-logarithmic scale in \textbf{b}. \textbf{c} Transfer curves of the Au contacted device showing the characteristics of an enhancement type, \textit{p}-channel semiconductor with a forward bias of V$_{\scriptsize \textrm{ds}}$ = +20 V. The semi-logarithmic scale in \textbf{d} shows a constant current in the subthreshold region.}
\label{fig2}
\end{figure}

To gain further insights into the response characteristics of the devices, additional electrical characterisation was carried out in the temperature range from 80 K to 40 K. These measurements were performed at decreasing temperatures to avoid thermal emission of captured charge carriers which would significantly alter the barrier height. The source-drain current \textit{vs} source-drain bias (I$_{\scriptsize \textrm{ds}}$-V$_{\scriptsize \textrm{ds}}$) response curves presented in figure \ref{fig2}a were measured on a device bearing Au contacts, using an applied gate bias (V$_{\scriptsize \textrm{gs}}$) of -40 V, which is in the devices ``open'' state.  It is apparent that the device exhibits a highly rectifying diode-like behaviour, which persists at lower temperatures. A graph on a semi-logarithmic scale, plotted in figure \ref{fig2}b, further emphasises the saturation of the source-drain current when the device is in reverse bias with respect to the source barrier. Other devices displayed similar trends and are shown in supplementary figure S1 online.

Fixing the source-drain bias to  a ``forward bias'' configuration (V$_{\scriptsize \textrm{ds}}=+20$ V), the source-drain current \textit{vs} gate bias (I$_{\scriptsize \textrm{ds}}$-V$_{\scriptsize \textrm{gs}}$) transfer curves were measured, and are plotted in figure \ref{fig2}c.  When the device is forward biased, it exhibits the characteristics of an enhancement type, \textit{p}-channel transistor with its threshold voltage (V$_{\scriptsize \textrm{th}}$) growing increasingly negative with decreasing temperature.  The semi-logarithmic scale plot, shown in figure \ref{fig2}d, shows the subthreshold region where the FET channel switches from its ``closed'' state to its ``open'' state and enters the linear regime.  At 80 K, the two-terminal hole mobility extracted from the linear region of the transfer graph is around 0.1 cm$^2$ V$^{-1}$ s$^{-1}$, which is in agreement with our previously published devices,\cite{Amit2017} but lower than the values reported by other groups.\cite{Nakaharai} We attribute the lower mobility to a high density of interface states, present either at the contacts or across the semiconducting flake. As we will show, this high density of traps contributes to the pinning of the Fermi level at the contacts and is instrumental in the realisation of ultra-low Schottky barriers. Other devices discussed in this report showed similar trends, see supplementary figure S2 online.  Contrary to MoS$_2$,\cite{Fontana2013} in MoTe$_2$ p-type behaviour dominates the transistor characteristics among all the metals used, with the exception of Ti which induces an asymmetric ambipolar behaviour, with p-doping at V$_{\scriptsize \textrm{gs}}$ = 0 V.

\begin{figure}[htb!]
\includegraphics[width=\linewidth]{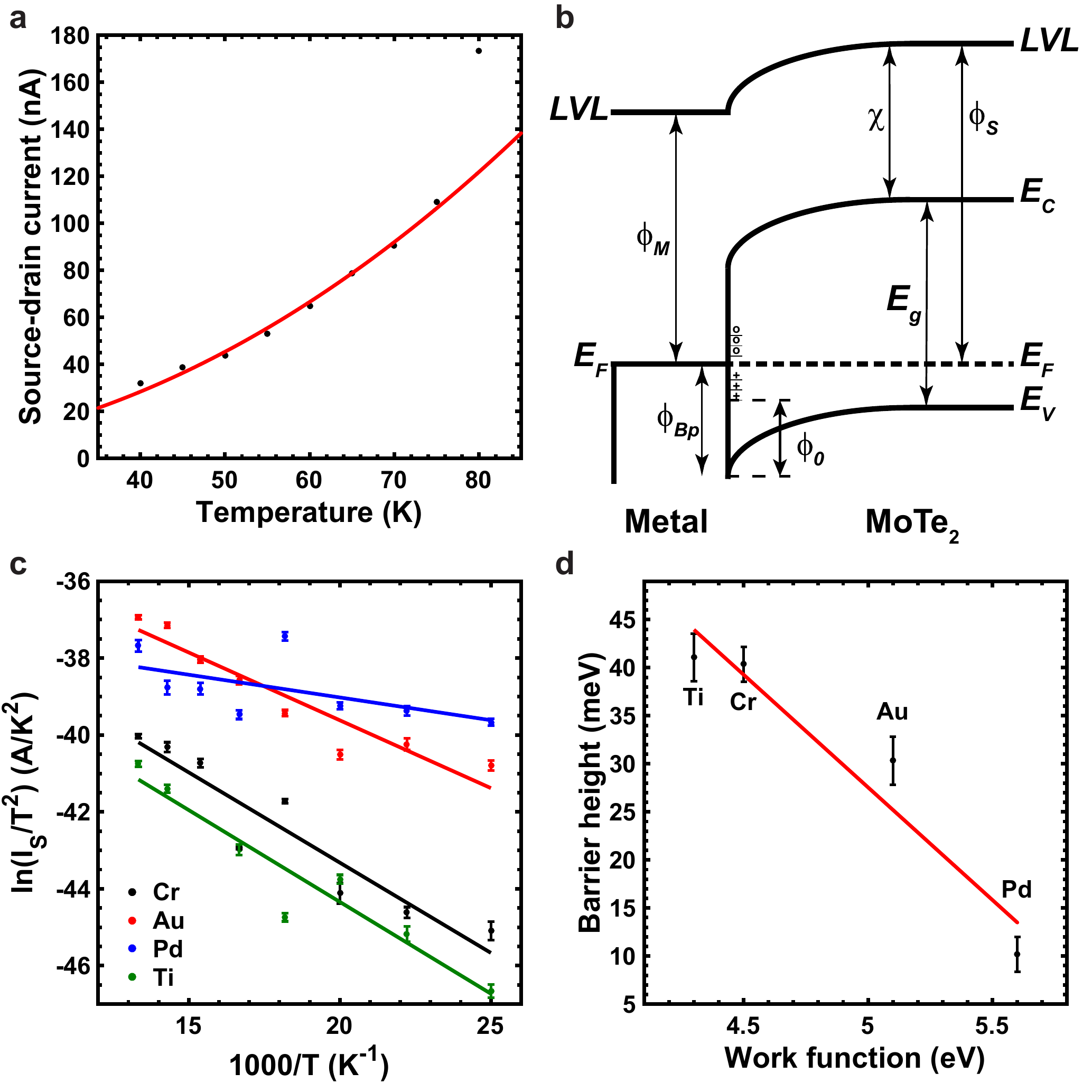}
\caption{\textbf{Barrier height determination}\\
\textbf{a} Current-temperature dependence for an Au contacted device in the ``open'' state, showing a good agreement with the thermionic emission model for temperatures below 80 K. \textbf{b} Schematic energy band diagram of the metal/\textit{p}-type semiconductor interface showing band bending of the conduction band edge $E_C$, the valence band edge $E_V$ and the local vacuum level $LVL$ when the Fermi levels $E_F$ are aligned and the system is in thermal equilibrium. The metal work function $\phi _M$, MoTe$_2$ work function $\phi _S$, electron affinity $\chi$, band gap $E_g$ and barrier height $\phi _{Bp}$ are noted in the diagram.  \textbf{c} A Richardson plot for the different devices at V$_{\scriptsize \textrm{gs}}-$V$_{\scriptsize \textrm{th}}=-11$ V. \textbf{d} The relation between barrier height and metal work function from which the extent of $E_F$ pinning can be determined.}
\label{fig3}
\end{figure}

The conductivity dependence on the temperature, determined by the response and transfer curves, which were acquired at different temperatures, reveals that the charge transport is dominated by a thermionic emission over an energy barrier. There are three main mechanisms of transport across an energy barrier: (1) thermionic emission from the high-end tail of Boltzmann distributed particles, (2) diffusion and (3) tunnelling (or field emission) through the barrier.\cite{Sze2007,Allain2015} For thermionic emission of charge carriers into a three-dimensional semiconductor, the current dependence with temperature ($T$) is given by the Schottky diode equation $I=AA^*T^2\exp(-E_A/kT)\left[\exp(qV/nk_BT)-1\right]$, where $A$ is the contact area, $A^*=4\pi m^* qk_B^2/h^3$ is the modified Richardson constant, $E_A$ is the activation energy required to overcome the barrier, $n$ is the ideality factor, $q$ is the basic charge and $k_B$ is the Boltzmann constant. Other mechanisms of charge transport differ mainly in the exponent of $T$, which is 0 for diffusion and 1 for tunnelling. To determine the transport mechanism, the MoTe$_2$ FET current \textit{vs.} temperature plot shown in figure \ref{fig3}a is fitted with I$_{\scriptsize \textrm{ds}} = B\cdot$ T$^{\alpha} e^{-C/T}$, where $\alpha$ is the exponent of T used as a fitting parameter and $B$ and $C$ are arbitrary constants. The use of the pre-exponential constant is justified by plotting the current at different temperatures with fixed V$_{\scriptsize \textrm{ds}}$ and fixed V$_{\scriptsize \textrm{gs}}$. The extracted $\alpha=2.082\pm0.211$ is in good agreement with the temperature exponent for thermionic emission, suggesting that this is the dominant transport mechanism governing the transport in the device, while diffusion and tunnelling of charges play a negligible role in the temperature range between 40 K and 80 K.  This is further supported by previous work that has shown the channel is depleted of its majority carriers,\cite{Amit2017} leading to the formation of barriers as wide as 10 $\mu$m in MoTe$_2$,\cite{Wang2016a} a fact which renders the efficiency of the tunnelling process insignificant.  Further discussion on the additional transport mechanisms are included in supplementary information section III online.

For thermionic emission to occur at low temperatures the thermal width of the Fermi-Dirac distribution ($\Delta E$) has to be comparable to or lower than the barrier height.  $\Delta E$ for the majority (80\%) of charge carriers can be found using the Fermi-Dirac distribution ($f(E)=1/(1+\exp((E-E_F)/k_B T))$)\cite{Sze2007} by considering the region between $f=0.1$ and $f=0.9$.  Within this range, the thermal distribution width is given by $\Delta E=4.4k_B T$.  Supplementary figure S6 online shows the thermal distribution width as a function of temperature.  Between the temperatures of  40 K and 80 K, the thermal distribution width ranges from 15 meV to 30 meV.  When the thermal energy distribution grows larger than the barrier height, additional mechanisms of charge transfer over an energy barrier, such as tunnelling currents and diffusion, become more prominent, rendering the thermionic emission mechanism irrelevant at higher temperatures.

As previously discussed, the thermionic emission occurs over the source/channel energy barrier that forms when the Fermi energies of the semiconductor and metal reach thermal equilibrium. The surface potential of the metal depletes the adjacent segment of the channel, effectively bending the energy bands of the semiconductor, as shown in figure \ref{fig3}b, forming the Schottky barrier. In the ideal case the SBH ($\phi _{Bp}$) is a function of the metal work function ($\phi _M$), the semiconductor electron affinity ($\chi$) and (for \textit{p}-type materials) the band gap ($E_g$), given by the Schottky-Mott rule, $\phi _{Bp}=E_g+\chi -\phi _M$.\cite{Guo2015b} To tune the barrier height, four different metals were used for the contacts, Ti, Cr, Au and Pd with work functions of 4.3\cite{Lin2014a}, 4.5,\cite{Attema2008a} 5.1\cite{Attema2008a} and 5.6 eV\cite{Singh-Miller2009} respectively. These work functions should theoretically form barrier heights of 480, 280, -320 and -820 meV respectively, based on an electron affinity of 3.78 eV\cite{Guo2016} and a band gap of 1.0 eV\cite{Goldstein2016a} for few-layer MoTe$_2$. Within this model a negative SBH is indicative of Ohmic contacts.  In a realistic interface, mid-gap states, that may be formed by vacancies,\cite{Guo2015b} oxidation,\cite{Nakaharai} and MIGS,\cite{Tung2014} increase the quantum capacitance of the band gap, an effect known as ``Fermi level pinning'', effectively altering the barrier height. In the extreme case, known as the Bardeen limit, a large density of mid-gap states can pin the barrier height completely.\cite{Guo2015b}

The MoTe$_2$ FET exhibits a diode-like behaviour, due to the pinned source/channel potential. As such, we may regard it as a single junction device, while treating the second junction as a fixed resistor. This point is further demonstrated in section II of the supplementary information online. In a single diode device, the height of the barrier can be extracted from temperature dependent transport measurements using a Richardson plot.\cite{Fathipour2014} To this end the response curves are fitted with the diode equation $I_{\scriptsize \textrm{ds}}=I_{\scriptsize \textrm{S}}\exp(-q(V_{\scriptsize \textrm{ds}}-I_{\scriptsize \textrm{ds}}R_{\scriptsize \textrm{S}})/nk_{\scriptsize \textrm{B}}T)- V_{\scriptsize \textrm{ds}}/R_{\scriptsize \textrm{P}}$, where $R_{\scriptsize \textrm{S}}$ is a series resistance that includes the channel resistance and the contact resistance and $R_{\scriptsize \textrm{P}}$ is a parallel shunt resistance. $I_{\scriptsize \textrm{S}}$ is the saturation current which is given by $I_{\scriptsize \textrm{S}}=AA^*T^2\exp(-\phi_{{\scriptsize \textrm{Bp}}}/k_{\scriptsize \textrm{B}}T)$ for thermionic emission. The extracted $I_{\scriptsize \textrm{S}}$ values are then plotted as $\ln(I_{\scriptsize \textrm{S}}/T^2)$ \textit{vs} $1/T$, shown in figure \ref{fig3}c and supplementary figure S7, which yield effective barrier heights of 41.1, 40.3, 30.3, and 10.2 meV for the Ti, Cr, Au and Pd contacts respectively.

The extraction of Schottky barrier heights as low as a few meVs can cause numerous difficulties due to competing mechanisms of transport.  At low temperatures, thermionic field emission can be the dominant mechanism for charge injection into the channel.\cite{Kenney2011}  However, our analysis of the source-drain current \textit{vs} temperature at fixed source-drain bias and gate bias in figure \ref{fig3}a clearly show that thermionic emission is the dominant mechanism.  In the determination of the barrier height, we used non-linear implicit function fitting tool to allow the physical parameters $I_S$, $R_S$, $R_P$ and $n$ to reach a global minimum of residuals in the parameters space.  For this, we have used a step-wise fit\cite{DeIacovo2015} to obtain the initial values of the parameters, which were then allowed to reach a higher degree of accuracy.  Indeed, the largest source of error in the fittings came from the noise level of the instruments with the small current signal at the lower temperatures.

It is important to note two major consequences of the measured results. First, while the resulting trend is in excellent qualitative agreement with the theoretical case where the barrier height decreases with an increase in metal work function, the extracted effective barrier heights differ profoundly from the calculated values. This indicates significant pinning of the Fermi level by mid-gap states. Second, the measured Schottky barrier heights are comparable with the Fermi Dirac thermal distribution width, thus confirming that thermionic emission over an energy barrier is the dominant current mechanism.

The extent of Fermi level pinning can be quantified by comparing the barrier height with the metal work function using $\phi_{\scriptsize \textrm{Bp}} = \phi_0 + S(\phi_{\scriptsize \textrm{M}}-\phi_0)$,\cite{Guo2015b,Kim} where $\phi_0$ is the reference energy of the mid-gap states. The pinning factor $S$ is equivalent to -1 in the ideal case with no pinning (Schottky limit) and 0 if $E_{\scriptsize \textrm{F}}$ is completely pinned (Bardeen limit). Figure \ref{fig3}d shows that the linear decrease in effective barrier height with increasing metal work function, has a gradient of -0.02 that indicates very strong pinning of the Fermi level. From the intercept of the linear fit, the reference energy of the mid-gap states is found to be 0.14 eV above the valence band maximum. The strong pinning also means that gate tunability of the effective Schottky barrier height is considerably suppressed in these devices. Indeed, only the Au and Pd contacted devices show some tunability, exhibiting SBH modulation in the order of up to 40 meV for gate bias shifts of 15 V. This effect is most likely due to image force barrier lowering, and is discussed in further detail in supplementary figure S8 and the following discussion online in supplementary information Sec. VI, however further work needs to be carried out to fully elucidate this mechanism.

Considering the low charge carrier mobility and the strong Fermi level pinning, we attribute the measured effective barrier heights to the high density of mid gap states present at the metal/semiconductor interface. We have previously shown that slow charge carrier dynamics in MoTe$_2$ strongly affects the performance of FETs.\cite{Amit2017} However, for photodetectors that are based on internal photoemission of charge carriers between the contact and the channel at that interface, the slow drift of charge carriers does not impede the performance of the detector, and can in fact serve as a short-lived reservoir for excited holes.

\section{Conclusion}
MoTe$_2$ FETs that exhibit ultra-low effective SBH were presented for the first time. The effective SBH can be controlled by changing the contact metal to some extent, but is dominated mainly by pinning of the Fermi level at the interface. The reference energy of the mid-gap states, which is a measure of the highest energy of occupied mid-gap states, was found to be 0.14 eV above the valence band maximum. The tunability of the barrier may assist in the design of internal photoemission based photodetectors, particularly for near-infrared applications.

\ack
NJT and SR acknowledge DSTL grant scheme Sensing and Navigation using quantum 2.0 technologies. IA acknowledges financial support from The European Commission Marie Curie Individual Fellowships (Grant number 701704). SR and MFC acknowledge financial support from EPSRC (Grant no. EP/J000396/1, EP/K017160/1, EP/K010050/1, EP/G036101/1, EP/M001024/1, EP/M002438/1), from Royal Society international Exchanges Scheme 2016/R1 and from The Leverhulme trust (grant title ”Quantum Drums” and ”Room temperature quantum electronics”).



\section*{Supplementary material}

*Nano* **11** 158896

*Supplementary Information*:

# Sub 20 meV Schottky barriers in metal/MoTe$_2$ junctions

Nicola J. Townsend[1], Iddo Amit[1], Monica F. Craciun[2] and Saverio Russo[1][†]

[1]Centre for Graphene Science, College of Engineering, Mathematics and Physical Sciences, University of Exeter, Exeter, EX4 4QL, U.K.

[2]Centre for Graphene Science, College of Engineering, Mathematics and Physical Sciences, University of Exeter, Exeter, EX4 4QL, U.K.

[†]`S.Russo@exeter.ac.uk`

# Contents

# I Response and Transfer curves of other devices

The response and transfer curves for devices bearing Ti, Cr and Pd metal contacts exhibit a similar trend to the Au contacted devices with the diode-like behaviour in the response curve, and an enhancement type *p*-channel FET in the transfer. The device bearing Ti metal contacts exhibits amipolar behaviour in the transfer curves with the smaller metal workfunction. The plots are shown in Supplementary Fig. S1 and S2, respectively. The charge carrier mobilities were extracted from the linear regions of the $I_{ds}$-$V_{gs}$ transfer curves at T = 80 K. and were found to be 0.01, 0.1 and 1.1 cm$^2$ V$^{-1}$ s$^{-1}$ for the Cr, Pd and Ti bearing devices respectively. These values are underestimated as two terminal measurements were used.

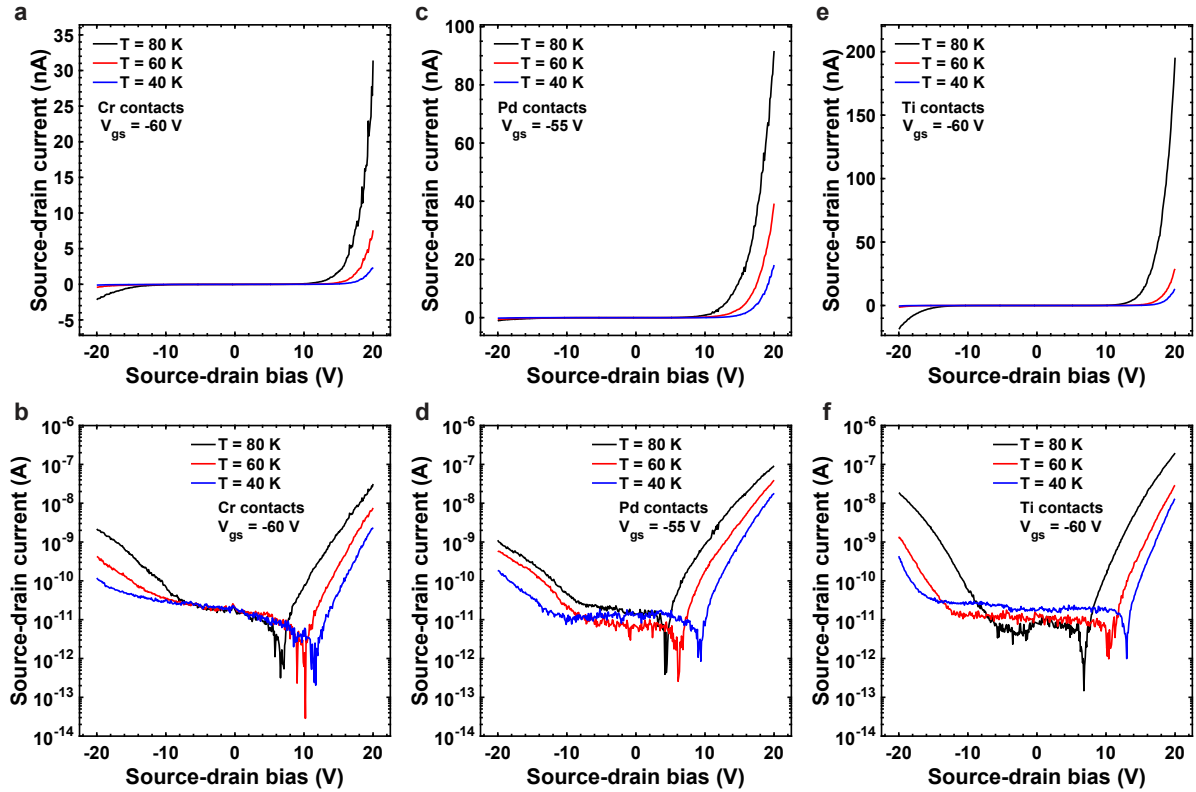

Figure S1: Response curves for metals with Cr (**a**,**b**), Pd (**c**,**d**), and Ti (**e**,**f**) contacts while the devices are in the "open" state. The top row (**a**,**c**,**e**) shows the plots on a linear scale and the bottom row (**b**,**d**,**f**) show the response curves on a semi-logarithmic scale.

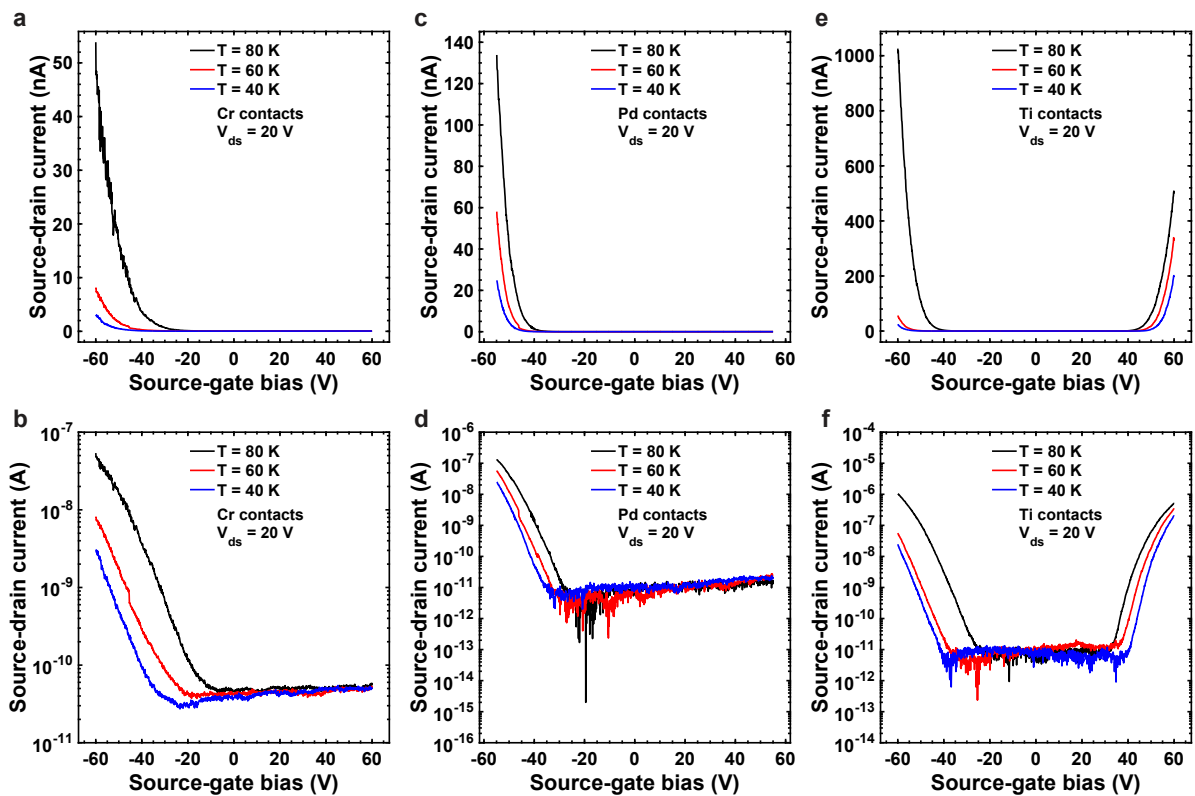

Figure S2: Transfer curves for FETs with Cr (**a,b**), Pd (**c,d**), and Ti (**e,f**) contacts while the devices are forward biased with $V_{ds}$ = +20V. The top row (**a,c,e**) shows the plots on a linear scale and the bottom row (**b,d,f**) show the response curves on a semi-logarithmic scale.

## II Pinning of source electrode

For the analysis shown in this report, the source electrode is assumed to be pinned at the contact to the extent that it can be considered as a constant resistor, and the gate modulation only affects the forward bias direction. This analysis is fundamentally different from the intuitive back-to-back diode architecture that should be present in the case of an unpinned interface. To justify our assumption of one "diode" being pinned we qualitatively compare our measured data to numerical models of a single diode and two back-to-back diode model in Supplementary Fig. S3.

The response curves at 80 K at different gate biases are shown in Supplementary Fig. S3a. To address the results with a correct model, two simplified cases were considered and solved numerically using the Trust-Region Dogleg algorithm (Newton Method) on Matlab. The first case is a single diode model[1] and was solved using a system of equations of the form:

$$I = I_S \left[ \exp\left( \frac{q(V_J - IR_s)}{nk_BT} \right) - 1 \right] \tag{S1a}$$

$$V_A = IR_s + V_J \tag{S1b}$$

Where $I$ is the current, $I_S$ is the saturation current, $q$ is the elementary charge, $V_J$ is the potential drop over the junction, $R_s$ is the series resistor that includes the "second" (non-dynamic) diode, $n$ is the ideality factor, $k_B$ is the Boltzmann constant, $T$ is the temperature and $V_A$ is the applied bias. To reflect the changes in junction conductivity with gate modulation, the value of $I_S$ in the model was reduced at high gate bias. This change is in accordance with the images charge barrier lowering model that is discussed in details in Supplementary Information Sec. VI.

The second case that was considered is the case of two back-to-back diodes,[1] where the device was modelled using the following system of equations:

$$I = I_S^{(1)} \left[ \exp\left( \frac{q(V_1 - IR_s)}{nk_BT} \right) - 1 \right] \tag{S2a}$$

$$I = I_S^{(2)} \left[ 1 - \exp\left( \frac{q(IR_s - V_2)}{nk_BT} \right) \right] \tag{S2b}$$

$$V_A = IR_s + V_1 + V_2 \tag{S2c}$$

Where $I_S^{(i)}$ is the saturation (reverse) current of the $i^{\text{th}}$ ($i = 1,2$) junction, and $V_i$ is the voltage drop on the $i^{\text{th}}$ junction.

Comparing the experimental curve to the single diode model (Supplementary Fig. S3b) and the two diode model (Supplementary Fig. S3c), it is clear that the measured data agrees with the single-diode case.

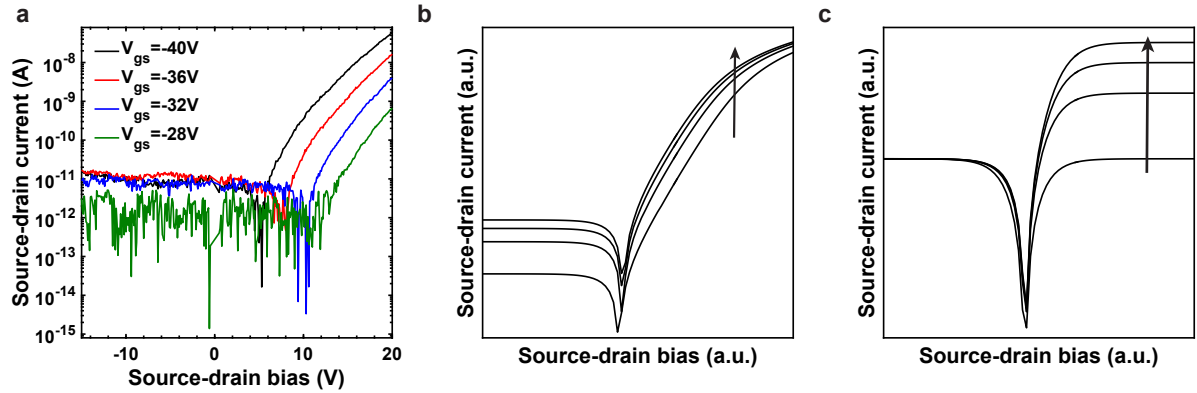

Figure S3: A comparison of the response curves of an Au bearing device at 80 K at various gate bias values (**a**) to calculated response curves of a single diode with a series resistor model (**b**) and those of two back-to-back diodes model (**c**).

## III   Transport mechanisms over an energy barrier

When injecting a current over an energy barrier, the three main injection mechanisms differ in their dependence on the temperature. For diffusion, the saturation current dependence on the temperature ($T$) is $I_D \propto \exp(-q\phi_{Bp}/k_B T)$, where $q$ is the elementary charge, $\phi_{Bp}$ is the barrier height, and $k_B$ is Boltzmann's constant. For field emission (tunnelling), the current is given by $I_{FE} \propto T \exp(-q(\phi_{Bp} - V_{ds})/E_{00})/\sin(\pi c_1 k_B T)$, where $c_1 = (1/2E_{00})\log(4(\phi_{Bp} - V_{ds})/\phi_p))$, $E_{00} = q\hbar\sqrt{N/4m^*\varepsilon_s}$, $\hbar$ is the reduced Planck's constant, $m^*$ is the effective mass of the charge carrier, and $N$ is the doping concentration. Finally, for thermionic emission, $I_S \propto T^2 \exp(-\phi_{Bp}/k_B T)$. From the three models, the difference in temperature dependence for each of the mechanisms arises from the exponent of the temperature in the pre-exponential factor.[1]

The fitting curves for devices bearing Cr, Pd and Ti contacts are presented in Supplementary Fig. S4 and found to be in good agreement with the thermionic emission model, up to 80 K.

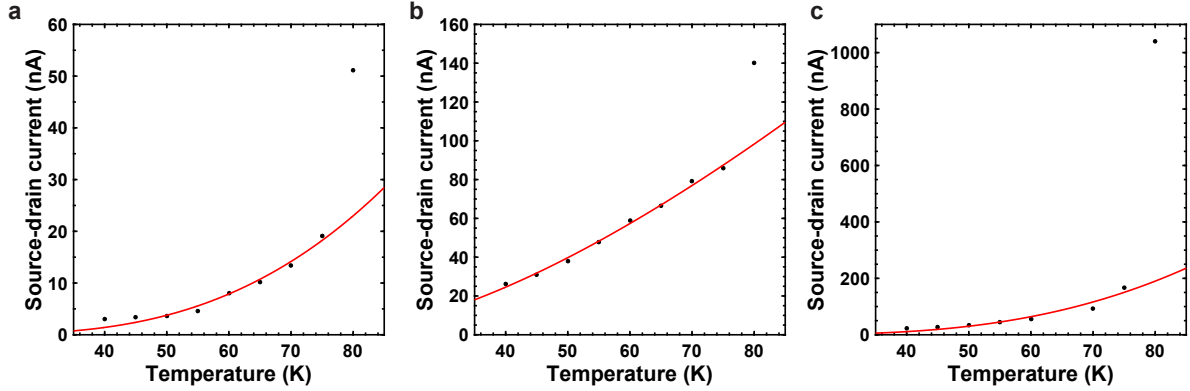

Figure S4: Current *vs.* temperature fittings for Cr (**a**), Pd (**b**) and Ti (**c**) with fits showing $\alpha_{Cr} = 2.571$, $\alpha_{Pd} = 2.081$ and $\alpha_{Ti} = 2.281$ at temperatures below 80 K.

For the device bearing Au contacts, the temperature dependence on source-drain current was also examined at different gate biases within the linear regime of the $I_{ds} - V_{gs}$ transfer curves, shown in Supplementary Fig. S5a. The fittings showed the temperature exponent of approximately 2 for the three values of $V_{gs}$, indicating thermionic emission is still the dominant mechanism.

For a different device bearing Pd contacts, these measurements were repeated over a wider temperature

range of 80 K to 300 K and is shown in Supplementary Fig. S5b. In this case, the fitting curves exhibited a mixed dependence on thermionic emission and tunnelling with $\alpha = 1.5424$, suggesting tunnelling contributions become significant at higher temperatures. As discussed in the main text and in more detail in Supplementary Information Sec. IV, this is in agreement with the thermal width of the charge carriers being greater than the measured barrier height, and therefore allowing tunnelling to occur. Also, with tunnelling being a significant contributor to transport over the barrier, thermionic emission in this temperature range cannot be solely used to determine barrier height, and is outside the scope of this work.

A qualitative measure of the contribution from each of the transport methods can be determined by comparing $k_B T$ to $E_{00}$. When $E_{00} << k_B T$, thermionic emission (TE) is dominant, whilst tunnelling (FE) is the main contributor to transport through the barrier when $E_{00} >> k_B T$. If $E_{00} \approx k_B T$, a combination of both thermionic emission and tunnelling (TFE) occurs. This is illustrated in Supplementary Fig. S5c, where theoretical values for MoTe$_2$ of $m^* = 0.3 m_e$ and $\varepsilon = 8$ were used.[2] This shows that for tunnelling to be a significant contributor to the transport across the barrier, the MoTe$_2$ would need to be highly doped to a carrier concentration above $10^{18}$ cm$^{-3}$, which is unlikely based on the presence of a high density of charge traps within the channel.[3] Therefore, at temperatures below 80 K, fitting of thermionic emission to $I_{ds} - V_{ds}$ response curves can be used to extract the barrier height.[4] However, further work needs to be carried out to quantise the doping concentration in MoTe$_2$, which is outside the scope of this paper.

Based on this model, the doping concentration should have increased to allow the significant contribution from tunnelling observed at temperatures between 80 K and 300 K. At the higher temperatures, the Fermi-Dirac distribution shows the thermal width becomes greater than the barrier height, as discussed in further detail in Supplementary Information Sec. IV. This would then cause tunnelling to become more important. However, further investigations need to be carried out to explain the increase in tunnelling at higher temperatures.

The alignment of the Fermi level across the metal/semiconductor at thermal equilibrium results in band bending on the semiconductor side. The lateral width over which the energy bands are bent, called the depletion width $W_D$, has a significant effect on the tunnelling probability, which decreases exponentially with $W_D$, as is shown schematically in Fig. S6. At low temperatures, the charge carriers have low energy with a narrow

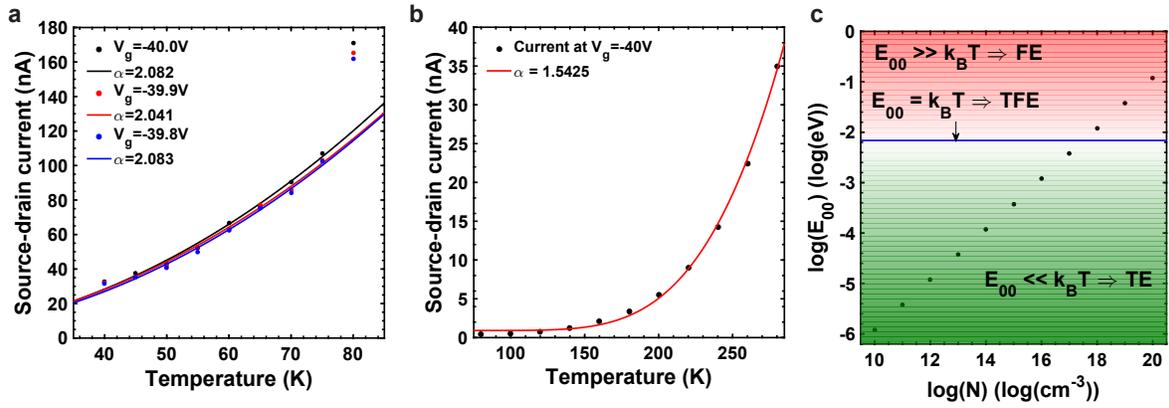

Figure S5: **a** Current *vs* temperature fittings for Au at different gate biases within the linear regime showing $\alpha_{V_g=-40.0V} = 2.082$, $\alpha_{V_g=-39.9V} = 2.041$ and $\alpha_{V_g=-39.8V} = 2.083$ at temperatures below 80 K. **b** Current *vs* temperature fittings for Pd over a wider temperature range with a fit showing $\alpha_{highT} = 1.5425$, suggesting tunnelling contributions become more significant at higher temperatures. **c** Relative contributions of thermionic emission (TE) when $E_{00} << k_BT$, tunnelling (FE) in the region $E_{00} >> k_BT$ and a combination of both (TFE) when $E_{00} \approx k_BT$.

distribution, resulting in a wide effective barrier which significantly reduces the probability of tunnelling is smaller than at higher temperatures. In the case of MoTe$_2$, the depletion width has been calculated to be of the order of 10 $\mu$m,[5] practically nullifying the probability for tunnelling. However, since the barrier height requires an energy which is comparable to the Fermi-Dirac thermal spread, charge injection into the channel can still occur by thermionic emission over the barrier.

## IV Fermi-Dirac Distribution curves

The Fermi-Dirac distribution shows the occupancy of states at energy $E$, as a function of temperature $T$ and the Fermi energy $E_F$:[1]

$$F_{FD}(E) = \frac{1}{1+\exp[(E-E_F)/k_BT]}, \quad (S3)$$

A plot of the Fermi-Dirac distribution at an arbitrary finite temperature is shown in Supplementary Fig. S7a.

Considering the majority (80 %) of charge carriers that are available for thermionic emission, we need

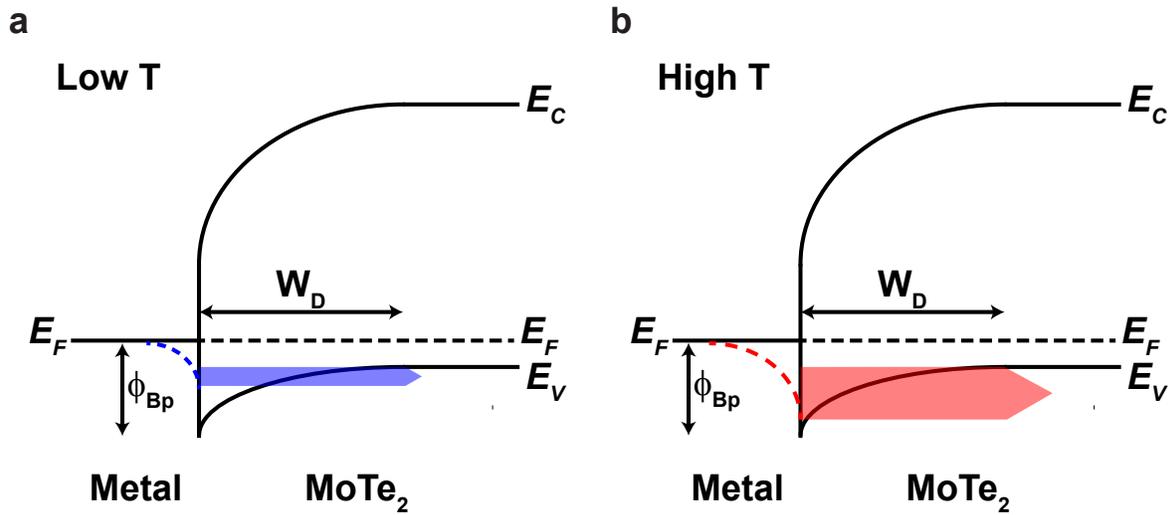

Figure S6: Schematic of the band bending caused by the formation of a Schottky barrier at the interface between a metal contact and the MoTe$_2$ channel showing the depletion width. At low temperatures (**a**) the depletion width associated with the low energy of the charge carriers is longer than that for the high temperature case (**b**) and, therefore, the tunnelling efficiency is lower.

to determine the thermal width $\Delta E$ which is in the range of energies bounded by $f_1 = F_{FD}(E_1) = 0.1$ and $f_2 = F_{FD}(E_2) = 0.9$. The charge carriers in this range are given by:[1]

$$\Delta E = kT \ln\left(\frac{0.81}{0.01}\right) \simeq 4.4kT \tag{S4}$$

The thermal width is linear with temperature and is plotted in Supplementary Fig. S7b, accounting for only thermally excited particles ($E < E_F$ for holes) it is clear that the thermal distribution width is of the same order of magnitude as the measured SBHs, supporting our observation that the charge injection mechanism is indeed thermionic emission, at the relevant temperatures ($< 80$ K).

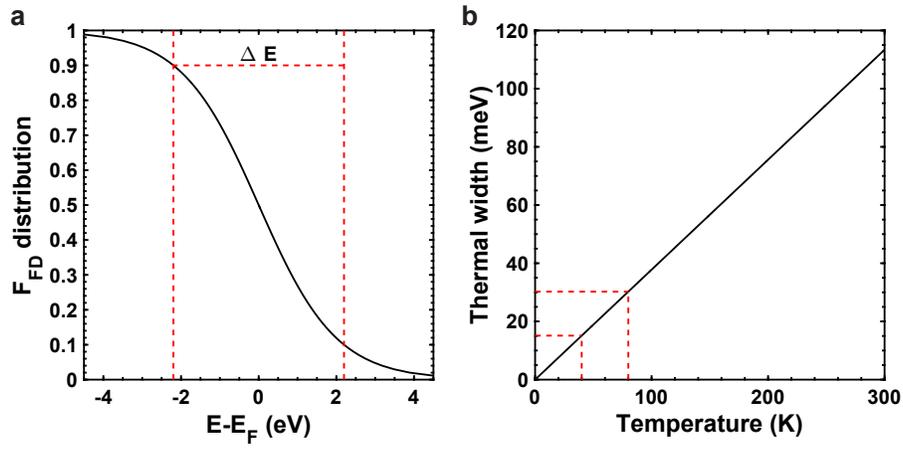

Figure S7: The Fermi-Dirac distribution at an arbitrary finite temperature (**a**) with the majority (80%) of charge carriers are in the range $f_{FD} = 0.1$ and $f_{FD} = 0.9$ shown by the red dashed line. This thermal width follows the temperature linearly as is shown in (**b**).

## V  Barrier height for each device at different gate biases

Further experimental results showing the Richardson plot for tested devices with all types of metal contacts used in this work are shown in Supplementary Fig. S8

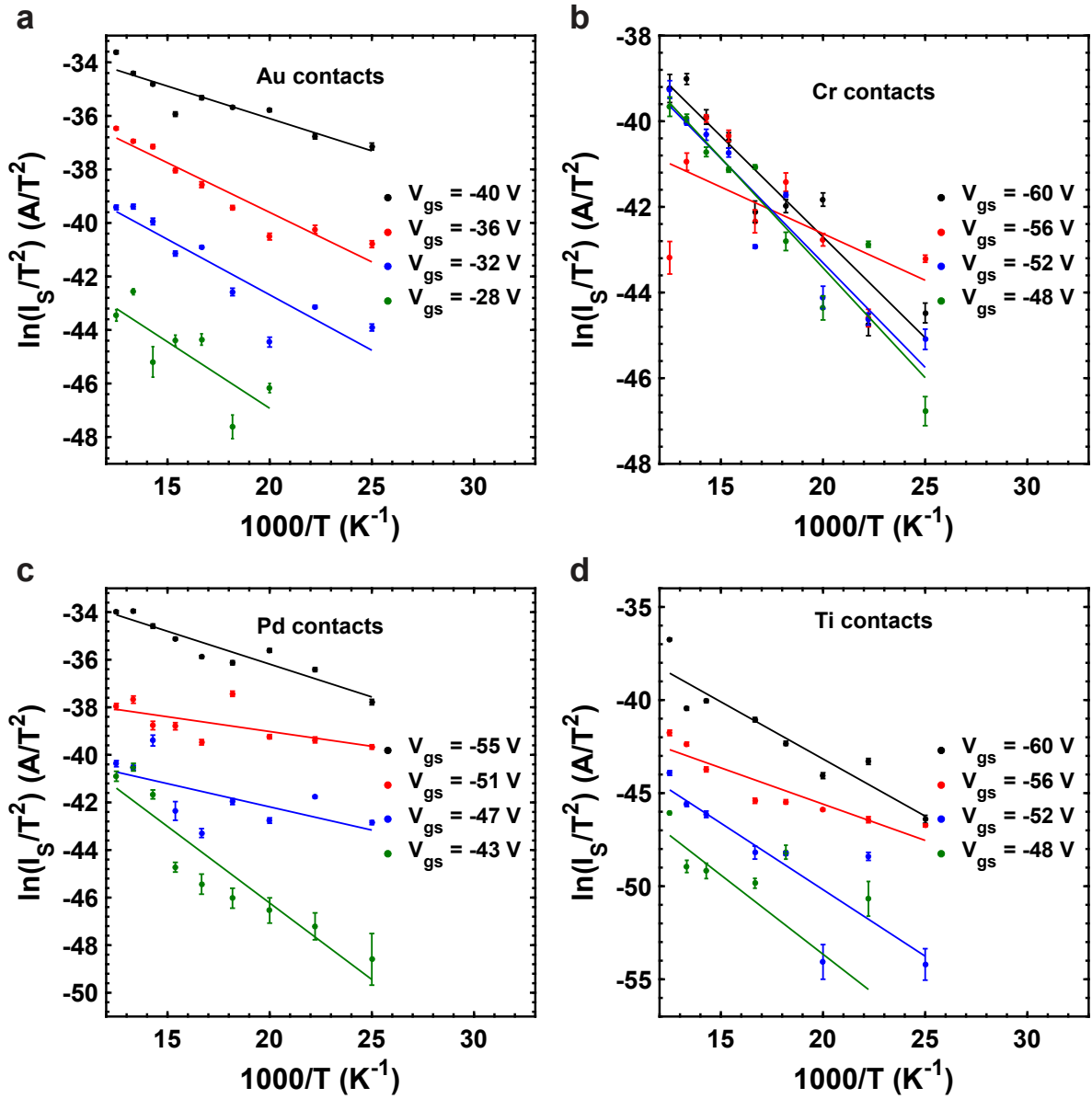

Figure S8: Richardson plots for the various gate voltages for Au (**a**), Cr (**b**), Pd (**c**) and Ti (**d**) showing similar trends.

## VI  Gate tuneability of Schottky barriers

Gate modulation of the effective barrier height was observed for the Au and Pd contacted devices, as shown in Supplementary Fig. S9a. This tuneability is most likely due to image force barrier lowering[1] where the SBH is reduced from its equilibrium value $\phi_{Bp0}$ by $\Delta\phi$ in the presence of an electric field, such as that provided by the gate bias. By shifting the Fermi level towards the valence band, the depletion region becomes smaller as

11

more free charge carriers are accumulated in the MoTe$_2$ channel. However, the voltage drop over the junction remains constant, resulting in a larger electric field over the depletion region which reduces the effective barrier height. There was no gate modulation observed for the Cr and Ti contacted devices, shown in Supplementary Fig. S9b. Supplementary Fig. S9c shows schematically two band diagrams, one for the a low gate bias (solid) and one for a higher bias value (dashed). These diagrams demonstrate the effect that a shift of the Fermi level has the depletion width ($W_{D1}$ and $W_{D2}$), which in turn changes the effective SBH by lowering the equilibrium barrier by $\Delta\phi_i$.

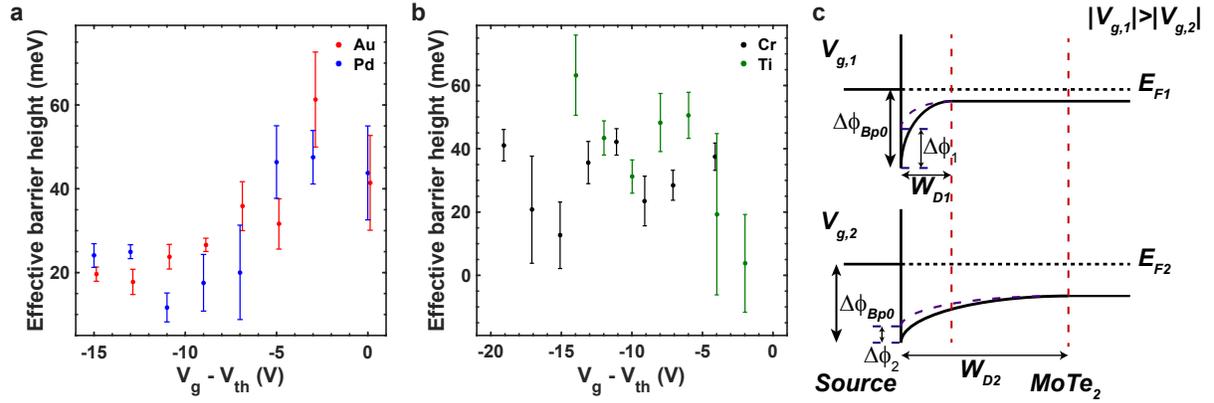

Figure S9: The effective Schottky barrier height at different gate biases for Au and Pd (**a**) shows a slight trend expected to be the result of image-force barrier lowering, as is shown in the schematic band diagram in **c**, whilst the effective barrier height appears to follow no such trend for the devices bearing Cr and Ti contacts (**b**).

# References

[1] Sze, S. M. & Ng, K. K. Physics of semiconductor devices. (Wiley, 2007).

[2] Ilatikhameneh, H. *et al.* Tunnel Field-Effect Transistors in 2-D Transition Metal Dichalcogenide Materials. *IEEE J. Explor. Solid-State Comput. Devices Circuits* **1,** 12-18 (2015).

[3] Amit, I. *et al.* Role of Charge Traps in the Performance of Atomically Thin Transistors. *Adv. Mater.* **29,** 1605598 (2017).




\section*{References}
\begin{thebibliography}{11}
\bibitem{wang2012electronics} Wang Q H, Kalantar-Zadeh K, Kis A, Coleman J N and Strano M S 2012 Electronics and optoelectronics of two-dimensional transition metal dichalcogenides \textit{Nat. Nanotechnol.} \textbf{7} 699–712
\bibitem{jariwala2014emerging} Jariwala D, Sangwan V K, Lauhon L J, Marks T J and Hersam M C 2014 Emerging device applications for semiconducting two-dimensional transition metal dichalcogenides \textit{ACS Nano} \textbf{8} 1102–20
\bibitem{Splendiani2010} Splendiani A, Sun L, Zhang Y, Li T, Kim J, Chim C-Y, Galli G and Wang F 2010 Emerging photoluminescence in monolayer MoS$_2$ \textit{Nano Lett.} \textbf{10} 1271–5
\bibitem{craciun2009trilayer} Craciun M F, Russo S, Yamamoto M, Oostinga J B, Morpurgo A F and Tarucha S 2009 Trilayer graphene is a semimetal with a gate-tunable band overlap \textit{Nat. Nanotechnol.} \textbf{4} 383–8
\bibitem{Cho2015} Cho S, Kim S, Kim J H, Zhao J, Seok J, Keum D H, Baik J, Choe D-H, Chang K J, Suenaga K, Kim S W, Lee Y H and Yang H 2015 Phase patterning for ohmic homojunction contact in MoTe$2$ \textit{Science} \textbf{349} 625–8
\bibitem{kang2013band} Kang J, Tongay S, Zhou J, Li J and Wu J 2013 Band offsets and heterostructures of two-dimensional semiconductors \textit{Appl. Phys. Lett.} \textbf{102} 12111
\bibitem{gong2013band} Gong C, Zhang H, Wang W, Colombo L, Wallace R M and Cho K 2013 Band alignment of two-dimensional transition metal dichalcogenides: Application in tunnel field effect transistors \textit{Appl. Phys. Lett.} \textbf{103} 53513
\bibitem{vabbina2015highly} Vabbina P, Choudhary N, Chowdhury A-A, Sinha R, Karabiyik M, Das S, Choi W and Pala N 2015 Highly sensitive wide bandwidth photodetector based on internal photoemission in CVD grown p-type MoS$_2$/Graphene Schottky junction \textit{ACS Appl. Mater. Interfaces} \textbf{7} 15206–13
\bibitem{Lao2014} Lao Y-F, Perera A G U, Li L H, Khanna S P, Linfield E H and Liu H C 2014 Tunable hot-carrier photodetection beyond the bandgap spectral limit \textit{Nat. Photonics} \textbf{8} 412–8
\bibitem{Guo2015b} Guo Y, Liu D and Robertson J 2015 3D behavior of Schottky barriers of 2D transition-metal dichalcogenides \textit{ACS Appl. Mater. Interfaces} \textbf{7} 25709–15
\bibitem{Allain2015} Allain A, Kang J, Banerjee K and Kis A 2015 Electrical contacts to two-dimensional semiconductors \textit{Nat. Mater.} \textbf{14} 1195–205
\bibitem{Kwon2017} Kwon J, Lee J-Y, Yu Y-J, Lee C-H, Cui X, Hone J and Lee G-H 2017 Thickness-dependent Schottky barrier height of MoS$_2$ field-effect transistors \textit{Nanoscale} \textbf{9} 6151–7
\bibitem{Lin2014a} Lin Y-F, Xu Y, Wang S-T, Li S-L, Yamamoto M, Aparecido-Ferreira A, Li W, Sun H, Nakaharai S, Jian W-B, Ueno K and Tsukagoshi K 2014 Ambipolar MoTe$_2$ transistors and their applications in logic circuits \textit{Adv. Mater.} \textbf{26} 3263–9
\bibitem{Pradhan2014} Pradhan N R, Rhodes D, Feng S, Xin Y, Memaran S, Moon B, Terrones H, Terrones M and Balicas L 2014 Field-effect transistors based on few-layered $\alpha$-MoTe$_2$ \textit{ACS Nano} \textbf{8} 5911–20
\bibitem{Fathipour2014} Fathipour S, Ma N, Hwang W S, Protasenko V, Vishwanath S, Xing H G, Xu H, Jena D, Appenzeller J and Seabaugh A 2014 Exfoliated multilayer MoTe$_2$ field-effect transistors \textit{Appl. Phys. Lett.} \textbf{105} 192101
\bibitem{Qi2017} Qi D, Wang Q, Han C, Jiang J, Zheng Y, Chen W, Zhang W and Wee A T S 2017 Reducing the Schottky barrier between few-layer MoTe$_2$ and gold \textit{2D Mater.} \textbf{4} 45016
\bibitem{Dawson1987a} Dawson W G and Bullett D W 1987 Electronic structure and crystallography of MoTe$_2$ and WTe$_2$ \textit{J. Phys. C Solid State Phys.} \textbf{20} 6159–74
\bibitem{Ruppert2014} Ruppert C, Aslan O B and Heinz T F 2014 Optical properties and band gap of single- and few-layer MoTe$_2$ crystals \textit{Nano Lett.} \textbf{14} 6231–6
\bibitem{Kan2015} Kan M, Nam H G, Lee Y H and Sun Q 2015 Phase stability and Raman vibration of the molybdenum ditelluride (MoTe$_2$) monolayer \textit{Phys. Chem. Chem. Phys.} \textbf{17} 14866–71
\bibitem{Chang} Chang Y, Lin C-Y, Lin Y and Tsukagoshi K 2016 Two-dimensional MoTe$_2$ materials: From synthesis, identification, and charge transport to electronics applications \textit{Jpn. J. Appl. Phys.} \textbf{55} 1102A1
\bibitem{Yamamoto2014} Yamamoto M, Wang S T, Ni M, Lin Y-F, Li S-L, Aikawa S, Jian W-B, Ueno K, Wakabayashi K and Tsukagoshi K 2014 Strong enhancement of Raman scattering from a bulk-inactive vibrational mode in few-layer MoTe$_2$ \textit{ACS Nano} \textbf{8} 3895–903
\bibitem{Thomsen2016a} Joshi J, Stone I R, Beams R, Krylyuk S, Kalish I, Davydov A V. and Vora P M 2016 Phonon anharmonicity in bulk T$_d$-MoTe$_2$ \textit{Appl. Phys. Lett.} \textbf{109} 31903
\bibitem{Tian2014} Tian H, Tan Z, Wu C, Wang X, Mohammad M A, Xie D, Yang Y, Wang J, Li L-J, Xu J and Ren T-L 2015 Novel field-effect Schottky barrier transistors based on Graphene-MoS$_2$ heterojunctions \textit{Sci. Rep.} \textbf{4} 5951
\bibitem{Amit2017} Amit I, Octon T J, Townsend N J, Reale F, Wright C D, Mattevi C, Craciun M F and Russo S 2017 Role of charge traps in the performance of atomically thin transistors \textit{Adv. Mater.} \textbf{29} 1605598
\bibitem{Nakaharai} Nakaharai S, Yamamoto M, Ueno K and Tsukagoshi K 2016 Carrier polarity control in $\alpha$-MoTe$_2$ Schottky junctions based on weak Fermi-level pinning \textit{ACS Appl. Mater. Interfaces} \textbf{8} 14732–9
\bibitem{Fontana2013} Fontana M, Deppe T, Boyd A K, Rinzan M, Liu A Y, Paranjape M and Barbara P 2013 Electron-hole transport and photovoltaic effect in gated MoS$_2$ Schottky junctions \textit{Sci. Rep.} \textbf{3} 1634
\bibitem{Sze2007} Sze S M and Ng K K 2007 Physics of semiconductor devices (Wiley)
\bibitem{Wang2016a} Wang F, Yin L, Wang Z, Xu K, Wang F, Shifa T A, Huang Y, Wen Y, Jiang C and He J 2016 Strong electrically tunable MoTe$_2$/graphene van der Waals heterostructures for high-performance electronic and optoelectronic devices \textit{Appl. Phys. Lett.} \textbf{109} 193111
\bibitem{Attema2008a} Attema J J, Uijttewaal M A, de Wijs G A and de Groot R A 2008 Work function anisotropy and surface stability of half-metallic CrO$_2$ \textit{Phys. Rev. B} \textbf{77} 165109
\bibitem{Singh-Miller2009} Singh-Miller N E and Marzari N 2009 Surface energies, work functions, and surface relaxations of low-index metallic surfaces from first principles \textit{Phys. Rev. B} \textbf{80} 235407
\bibitem{Guo2016} Guo Y and Robertson J 2016 Band engineering in transition metal dichalcogenides: Stacked versus lateral heterostructures \textit{Appl. Phys. Lett.} \textbf{108} 233104
\bibitem{Goldstein2016a} Goldstein T, Chen S, Tong J, Xiao D, Ramasubramaniam A and Yan J 2016 Raman scattering and anomalous Stokes-anti-Stokes ratio in MoTe$_2$ atomic layers \textit{Sci. Rep.} \textbf{6} 28024
\bibitem{Tung2014} Tung R T 2014 The physics and chemistry of the Schottky barrier height \textit{Appl. Phys. Rev.} \textbf{1} 11304
\bibitem{Kenney2011} Kenney C, Saraswat K C, Taylor B and Majhi P 2011 Thermionic Field Emission Explanation for Nonlinear Richardson Plots \textit{IEEE Trans. Electron Devices} \textbf{58} 2423–9
\bibitem{DeIacovo2015} De Iacovo A, Colace L, Assanto G, Maiolo L and Pecora A 2015 Extraction of Schottky Barrier Parameters for Metal–Semiconductor Junctions on High Resistivity Inhomogeneous, Semiconductors \textit{IEEE Trans. Electron Devices} \textbf{62} 465–70
\bibitem{Kim} Kim C, Moon I, Lee D, Choi M S, Ahmed F, Nam S, Cho Y, Shin H-J, Park S and Yoo W J 2017 Fermi level pinning at electrical metal contacts of monolayer molybdenum dichalcogenides \textit{ACS Nano} \textbf{11} 1588–96
\end{thebibliography}
\end{document}